\documentclass{article}
\usepackage{amsmath}
\newtheorem{definition}{Definition}
\newtheorem{assumption}{Physical Assumption}
\newtheorem{theorem}{Theorem}

\begin{document}

\title{A tensor interpretation of the 2D Dirac equation}
\author{Dmitri Vassiliev
\thanks{
Department of Mathematical Sciences, University of Bath, Bath BA2 7AY,
United Kingdom;
D.Vassiliev@bath.ac.uk;
http://www.bath.ac.uk/\~{}masdv/;
research supported by a Leverhulme Fellowship.}
}
\maketitle

\begin{abstract}
We consider the Dirac equation in flat Minkowski 3--space and
rewrite it as the Maxwell equation in Minkowski 4--space
with torsion. The torsion tensor is defined as the dual
of the electromagnetic vector potential.
Our model clearly
distinguishes the electron and the positron without
resorting to ``negative frequencies'': we produce a
real scalar invariant (charge) which indicates whether we are
looking at an electron or a positron.
Another interesting feature of our model is that
the free electron and positron are identified with gradient type
solutions of the standard (torsion free) Maxwell equation;
such solutions have traditionally
been disregarded on the grounds of gauge invariance.
\end{abstract}

\section{Introduction}
\label{intro}

The Dirac equation is the following system of 4 partial differential
equations in Minkowski 4--space:
\begin{equation}
\label{Dirac3D}
\begin{pmatrix}
\ \ i\nabla_{\!0}-1\ \ &
\ \ 0\ \ &
\ \ i\nabla_{\!3}\ \ &
\ \ i\nabla_{\!1}+\nabla_{\!2}\\
\ \ 0\ \ &
\ \ i\nabla_{\!0}-1\ \ &
\ \ i\nabla_{\!1}-\nabla_{\!2}\ \ &
\ \ -i\nabla_{\!3}\\
\ \ -i\nabla_{\!3}\ \ &
\ \ -i\nabla_{\!1}-\nabla_{\!2}\ \ &
\ \ -i\nabla_{\!0}-1\ \ &
\ \ 0\\
\ \ -i\nabla_{\!1}+\nabla_{\!2}\ \ &
\ \ i\nabla_{\!3}\ \ &
\ \ 0\ \ &
\ \ -i\nabla_{\!0}-1
\end{pmatrix}
\begin{pmatrix}
\phi_1\\\phi_2\\\chi_1\\\chi_2
\end{pmatrix}
=0
\end{equation}
where
\begin{equation}
\label{complex1}
\nabla=\partial+iA,
\end{equation}
$\partial$ being the operator of partial differentiation
and $A$ the vector potential of the external electromagnetic
field (given real valued vector function). Equation
(\ref{Dirac3D}) is often referred to as the 3D (3--dimensional)
Dirac equation, with 3 indicating the number of spatial variables.

The set of complex quantities
$\psi=(\phi_1\ \phi_2\ \chi_1\ \chi_2)^T$
is a bispinor, and the way it behaves
under Lorentz transformations of coordinates
is quite extraordinary. For example,
a spatial rotation of the coordinate system by an angle $2\pi$
changes the sign of $\psi$. More important, time reversal
leads to complex conjugations resulting in the well known
difficulty of distinguishing the electron and the positron
(problem of ``negative frequencies''). See Sections 18 and 19
in \cite{LL4} for details.

Our aim is to provide a satisfactory tensor interpretation of
(\ref{Dirac3D}). We fail to achieve this goal in full,
but succeed in handling the
simpler case of the 2D Dirac equation. The 2D Dirac equation
is a special case of (\ref{Dirac3D}) arising
when $A$ and $\psi$ do not depend on $x^3$ and $A_3\equiv0$.

The essential new elements our mathematical model are as follows.

$\bullet$
The Dirac equation is interpreted as a perturbation
of the polarised Maxwell equation
\begin{equation}
\label{Maxwell1}
*du=\pm idu
\end{equation}
rather than an independent equation in its own right.

$\bullet$ The external field (perturbation)
is introduced into the model as
a classical real differential geometric connection with torsion
\begin{equation}
\label{torsion1}
T=*A
\end{equation}
rather than by means of the complex formula (\ref{complex1}).

$\bullet$ In our model the electron
(solution of the perturbed Maxwell equation)
is considered simultaneously
with the photon
(solution of the unperturbed Maxwell equation).
Algebraically this seems to be the only way of
avoiding the appearance of multiple copies of the Dirac equation
as in \cite{BT}.

$\bullet$ The electron mass is introduced into the model as a
prescribed oscillation
along the $x^3$ coordinate. Note that the idea of viewing mass
in terms of oscillation along an additional
space--like coordinate is a
classical one, going back to Oskar Klein \cite{K}.
The peculiarity of the
2D Dirac equation is that one can use the third spatial coordinate
for this purpose, thus avoiding the necessity of
introducing a fifth dimension.

The paper has the following structure.

In Section~\ref{notation} we specify our notation.
In Section~\ref{mass} we define the electron mass.
In Section~\ref{photon} we describe our mathematical model
for the photon.
Section~\ref{torsion} gives basic geometric facts
concerning Minkowski 4--space with torsion.
In Section~\ref{electron} we describe our mathematical model
for the electron/positron.
In Section~\ref{basic} we
identify a basic symmetry of our model
with respect to complex conjugation; in particular,
we explain why our tensor model
is free of the problem of ``negative frequencies''.
In Section~\ref{main} we state and prove the main result
of this paper, Theorem \ref{maintheorem};
this theorem establishes the
equivalence of our tensor model and the 2D Dirac equation.
In Section~\ref{free} we
introduce the notion of a free ($A\equiv0$) electron/positron.
Finally, in Section~\ref{charge}
we define a real scalar invariant (charge) which allows us
to distinguish the electron and positron solutions.

\section{Principal notation}
\label{notation}

We work in Minkowski 4--space equipped with coordinates
$(x^0,x^1,x^2,x^3)$
and metric $g_{\mu\nu}=\operatorname{diag}(+1,-1,-1,-1)$.
We use Greek letters for tensor indices, with the exception
of the letters $\alpha$ and $\beta$ which have a special meaning.
Tensor indices take the values 0, 1, 2, 3.
We denote $\partial_\mu=\partial/\partial x^\mu$.
Our system of units is such that the speed of light $c$,
Planck's constant $\hbar$, and the electron mass $m$ have the value 1.
The Dirac equation (\ref{Dirac3D}), (\ref{complex1}) is written
as in \cite{LL4}, Section~21 (standard representation).
The only difference is that we have incorporated the electron charge
into the vector potential: our $A$ corresponds to
the $eA$ of \cite{LL4}.

We work with complex valued antisymmetric tensor functions of
various ranks. All functions are assumed to be infinitely smooth.
We denote complex conjugation with an ``overline''.

Given a pair of antisymmetric tensors
$Q$ and $R$ of the same rank $q$
we denote
$Q\cdot R:=\frac1{q!}Q_{\mu_1\ldots\mu_q}R^{\mu_1\ldots\mu_q}$.
We write the condition $Q\cdot\overline R=0$ as $Q\perp R$.

Tensor functions of the form
\begin{equation}
\label{plane}
\text{constant tensor}
\ \times\ e^{-ik\cdot x},
\end{equation}
$k$ real,
are called plane waves. The vector $k$ is called the wave vector.
In defining a plane wave as
$\sim e^{-ik\cdot x}$ rather than $\sim e^{ik\cdot x}$
we follow the convention of \cite{LL2}, \cite{LL4}.

By $\varepsilon_{\kappa\lambda\mu\nu}$ we denote the totally
antisymmetric tensor.
We specify an orientation of our Minkowski 4--space, and put
\begin{equation}
\label{total1}
\varepsilon_{0123}:=+1
\end{equation}
for all coordinate systems with positive orientation.

We define the action of the Hodge star
(duality transformation) on an
antisymmetric tensor $Q$ of rank $q$ as
\begin{equation}
\label{Hodge1}
(*Q)_{\mu_{q+1}\ldots\mu_4}:=\frac1{q!}\,
Q^{\mu_1\ldots\mu_q}\varepsilon_{\mu_1\ldots\mu_4}.
\end{equation}

Let $Q$ and $R$ be antisymmetric tensors of rank $q$ and $r$,
respectively. We define their exterior product as
\begin{equation}
\label{extprod1}
(Q\wedge R)_{\lambda_1\ldots\lambda_{q+r}}:=
\frac1{q!r!}\sum\operatorname{sgn}(P)
Q_{\mu_1\ldots\mu_q}R_{\nu_1\ldots\nu_r}
\end{equation}
where summation is carried out over all permutations
$P=
\genfrac{(}{)}{0pt}{}
{\lambda_1\ldots\lambda_{q+r}}
{\mu_1\ldots\mu_q\nu_1\ldots\nu_r}$.

Let $Q$ and $R$ be antisymmetric tensors functions
of the same rank $q$. We define their inner product as
\begin{equation}
\label{innprod1}
(Q,R):=\int Q\cdot\overline R\ dx^0\,dx^1\,dx^2\,dx^3\,.
\end{equation}

We denote by
\begin{equation}
\label{extder1}
dQ:=\partial\wedge Q
\end{equation}
the exterior derivative and by $\delta$ its adjoint with respect
to the inner product (\ref{innprod1}).

Our definitions
(\ref{total1})--(\ref{extder1})
agree with those in \cite{N},
modulo the fact that we use the language of antisymmetric tensors
rather than that of differential forms.

Lorentz transformations are assumed to be ``passive'' in the sense
that we transform the coordinate system and not the tensors
themselves.

We assume that our Minkowski 4--space has a specified
coordinate axis $x^3$. This means that we only allow
Lorentz transformations which preserve the equation
of the hyperplane $\{x^3=0\}$.
Such Lorentz transformations are not necessa\-rily proper: an example
of an improper one is the reversal of the $x^3$ coordinate.

Given an antisymmetric tensor $Q$ we define another
antisymmetric tensor
\begin{equation}
\label{reflection}
({\mathcal R}Q)_{\mu_1\ldots\mu_q}:=
\begin{cases}
Q_{\mu_1\ldots\mu_q}\quad\text{if}\quad 3\not\in
\{\mu_1,\ldots,\mu_q\},
\\
-Q_{\mu_1\ldots\mu_q}\quad\text{if}\quad 3\in
\{\mu_1,\ldots,\mu_q\}.
\end{cases}
\end{equation}
The tensor ${\mathcal R}Q$ is the reflection of $Q$
about the the hyper\-plane $\{x^3=0\}$.
The ``active'' reflection operator ${\mathcal R}$
should not be confused
with the ``passive'' reversal of the $x^3$ coordinate.

\section{Mass}
\label{mass}

Throughout this paper we will be dealing with tensor functions
of the form
\begin{equation}
\label{mass1}
Q(x^0,x^1,x^2,x^3)=\tilde Q(x^0,x^1,x^2)e^{\pm ix^3}.
\end{equation}
Condition (\ref{mass1}) introduces a length scale into our model,
which we interpret as the Compton wave length of the electron.
In view of our choice of the system of units $c=\hbar=m=1$ we
can use (\ref{mass1}) as the definition of the electron mass.

In the next section we shall acquire a second set of
$\pm$ signs, independent of the one in (\ref{mass1}).
In order to avoid a clash of notation we shall write (\ref{mass1}) as
\begin{equation}
\label{mass2}
Q(x^0,x^1,x^2,x^3)=\tilde Q(x^0,x^1,x^2)e^{-i\alpha x^3}
\end{equation}
where the index $\alpha$ takes the values $\pm1$.
We put an extra minus in the right hand side of
(\ref{mass2}) because it is convenient in view of our definition
of a plane wave (\ref{plane}).

\section{Mathematical model for the photon}
\label{photon}

In the absence of sources the Maxwell equation in vector form is
\begin{equation}
\label{Maxwell2}
\delta du=0
\end{equation}
where $u$ is the unknown vector function. A solution $u$
is said to be polarised if the corresponding electromagnetic
tensor $du$ is an eigenvector of the linear operator $*$.
This polarisation condition is precisely formula (\ref{Maxwell1}).

Let us now view the polarisation condition
(\ref{Maxwell1}) as a differential equation and compare it with
(\ref{Maxwell2}).
Using the fact that $\delta*d=0$,
it is easy to see that $u$ is a polarised solution of
(\ref{Maxwell2}) if and only if it is a solution of
(\ref{Maxwell1}). Therefore we shall call (\ref{Maxwell1})
the polarised Maxwell equation.

In order to avoid a clash of notation we shall write
(\ref{Maxwell1}) as
\begin{equation}
\label{Maxwell3}
*du=i\beta du
\end{equation}
where the index $\beta$ takes the values $\pm1$.

Equation (\ref{Maxwell3}) is under-determined because
it is actually a system
of 3 equations with 4 unknowns. This under-determinacy does not
cause problems because (\ref{Maxwell3}) admits
an obvious gauge transformation:
if $u$ is a solution of (\ref{Maxwell3}) then so is
$u+ds$, where $s$ is an arbitrary scalar function.
One may find it convenient to complement
(\ref{Maxwell3}) by a gauge condition which would exclude
the possibility of adding an arbitrary gradient and bring the
total number of equations up to 4.
In our setting the natural gauge is
\begin{equation}
\label{gauge1}
u_3=0.
\end{equation}
The gauge condition (\ref{gauge1}) is perfectly
suited for our purposes: it totally
excludes the possibility of adding a gradient because
in view of (\ref{mass2})
\ $(ds)_3=-i\alpha s$, and the only way $(ds)_3$ can be zero
is if $s$ is zero. Nevertheless, in the following definition
we do not insist on a particular gauge. The reason for not
doing this will become clear later, when it will
emerge (see (\ref{pertgauge1}))
that our equation for the electron/positron does not depend
on the choice of the gauge for the photon.

We call a solution of (\ref{Maxwell3}) trivial if it is the gradient
of a scalar function.

\begin{definition}
\label{photon1}
A nontrivial plane wave solution $\ u\ $
of the under-determined equation
(\ref{Maxwell3}) is called a photon.
\end{definition}

\section{Minkowski space perturbed by torsion}
\label{torsion}

\subsection{Connection generated by the external field}
\label{connection}
Let us now equip our Minkowski 4--space
with a non-trivial connection.
This means that we will have to start distinguishing the usual
partial derivative $\partial$ and the covariant derivative
$\nabla$. When acting on a vector function $v$
the general formulae relating the two are
\begin{equation}
\label{connection1}
\nabla_{\!\mu}v^\lambda=\partial_\mu v^\lambda
+{\Gamma^\lambda}_{\mu\nu}v^\nu,
\qquad
\nabla_{\!\mu}v_\nu=\partial_\mu v_\nu
-{\Gamma^\lambda}_{\mu\nu}v_\lambda.
\end{equation}
Here the notation is from \cite{N}.

We take the connection coefficients to be
\begin{equation}
\label{connection2}
{\Gamma^\lambda}_{\mu\nu}=
\frac12A_\kappa{\varepsilon^{\kappa\lambda}}_{\mu\nu}
\end{equation}
where $A$ is the vector potential of the external electromagnetic
field.

\

\noindent{\it Remark 1\ }
In a general coordinate system the right hand side
of (\ref{connection2}) would have the Christoffel symbol
$
\genfrac{\{}{\}}{0pt}{}{\lambda}{\mu\nu}=
\frac12g^{\lambda\kappa}
(\partial_\mu g_{\nu\kappa}
+\partial_\nu g_{\mu\kappa}
-\partial_\kappa g_{\mu\nu})
$
as an additional term. We dropped it
because following the traditions of special relativity
we restrict ourselves to coordinate systems
in which the metric tensor is constant.

\

The connection coefficients (\ref{connection2}) satisfy
${\Gamma^\kappa}_{\lambda\mu}g_{\kappa\nu}
+{\Gamma^\kappa}_{\lambda\nu}g_{\kappa\mu}=0$
so our connection is metric compatible.

Defining the torsion tensor by the standard formula
${T^\lambda}_{\mu\nu}=
{\Gamma^\lambda}_{\mu\nu}-{\Gamma^\lambda}_{\nu\mu}$
we arrive at (\ref{torsion1}). Conversely, given torsion
(\ref{torsion1}) we uniquely recover
(see formula (7.34) in \cite{N})
the coefficients (\ref{connection2}) of our metric compatible
connection.

\subsection{Generalisation of the notion of exterior derivative}
\label{genext}
The natural generalisation of (\ref{extder1}) is the operator
\begin{equation}
\label{genext1}
d_{\!A}Q:=\nabla\wedge Q.
\end{equation}
For an antisymmetric tensor function $Q$ of rank $q$
formula (\ref{genext1}) is understood in the following way:
we write $\nabla\wedge Q$ in accordance with
(\ref{extprod1}) getting $(q+1)!$ terms of the type
$\nabla_\mu Q_{\nu_1\ldots\nu_q}$, and expand each of
these terms in accordance with the standard rules of covariant
differentiation of a rank $q$ tensor
(see formula (7.26) in \cite{N}).
In particular, when $Q=v$ is
a vector function we get, by applying (\ref{connection1}),
\begin{multline}
\label{genext2}
(d_{\!A}v)_{\mu\nu}=\nabla_\mu v_\nu-\nabla_\nu v_\mu
=\partial_\mu v_\nu-{\Gamma^\lambda}_{\mu\nu}v_\lambda
-\partial_\nu v_\mu+{\Gamma^\lambda}_{\nu\mu}v_\lambda
\\
=(dv)_{\mu\nu}-{T^\lambda}_{\mu\nu}v_\lambda.
\end{multline}
We see that the operator $d_{\!A}$ differs from
%the usual exterior derivative
$d$ by terms with torsion.

\

\noindent{\it Remark 2\ }
The author's impression is that in mathematics literature it is
not customary to work with the operator (\ref{genext1})
and to view it as a natural generalistion of (\ref{extder1}).
On the other hand, it appears that
in physics literature (\ref{genext1}) is
accepted as the natural way of forming a higher rank
antisymmetric tensor; see, for example,
\cite{LL2}, Section~90. For the vast majority of applications the
matter of distinguishing (\ref{extder1}) and (\ref{genext1}) is,
however, irrelevant, because these applications normally concern
Levi--Civita connections,
in which case
(\ref{extder1}) and (\ref{genext1}) define the same operator.

\

Using (\ref{torsion1}) we can rewrite (\ref{genext2}) in shorter
form as
\begin{equation}
\label{genext3}
d_{\!A}v=dv-*(A\wedge v).
\end{equation}
Formula (\ref{genext3}) will play a central role in
the proof of Theorem \ref{maintheorem}.

\section{Mathematical model for the electron/positron}
\label{electron}
The perturbed analogue of the polarised Maxwell equation
(\ref{Maxwell3}) is
\begin{equation}
\label{pertMax1}
*d_{\!A}v=i\beta d_{\!A}v.
\end{equation}

The crucial difference between (\ref{Maxwell3}) and (\ref{pertMax1})
is that the latter does not admit the usual gauge transformation
because for a scalar function $s$ we have
$d_{\!A}d_{\!A}s=d_{\!A}ds=-*(A\wedge ds)\ne0$.
Therefore the choice of a condition complementing (\ref{pertMax1})
becomes a matter of principle rather than a matter of convenience.

We fix an arbitrary photon $u$ and impose not one, but two
conditions
\begin{equation}
\label{pertgauge1}
v\perp u,\qquad v\perp k
\end{equation}
where $k$ is the wave vector of the photon. As
any two photons corresponding
to the same $k$ differ by a gradient, an equivalent way of
imposing the conditions (\ref{pertgauge1}) is to require
$v$ to be orthogonal to all photons $u$ with given wave vector.

We call a solution of (\ref{pertMax1}), (\ref{pertgauge1})
trivial if it is identically zero.

\begin{definition}
\label{electron1}
A nontrivial solution $\ v\ $ of the over-determined system
(\ref{pertMax1}), (\ref{pertgauge1}) is called an
electron/positron.
\end{definition}

\section{Basic symmetry}
\label{basic}

Before proceeding to the actual analysis of our tensor model
let us point out its basic symmetry:
if $u$, $v$ are solutions of
(\ref{Maxwell3}), (\ref{pertMax1}), (\ref{pertgauge1})
with indices $\alpha=\alpha_0$, $\beta=\beta_0$, then
$\overline u$, $\overline v$ are solutions of
(\ref{Maxwell3}), (\ref{pertMax1}), (\ref{pertgauge1})
with indices $\alpha=-\alpha_0$, $\beta=-\beta_0$. This is
obvious because $\,i\,$ comes into our model multiplied by
$\alpha$ (see (\ref{mass2})) or $\beta$
(see (\ref{Maxwell3}), (\ref{pertMax1})).
The argument relies on
the fact that we introduced the external field as a real connection
as opposed to
the traditional complex formula (\ref{complex1}).

As a consequence, our model is free of the problem of ``negative
frequencies''. Without loss of generality we shall assume further
on that the wave vector $k$ of the chosen photon $u$ lies on the forward
light cone, i.e., $k_0>0$.

\section{Main result}
\label{main}

Let us write down explicitly the 2D Dirac equation. In doing this we
should avoid using the notation (\ref{complex1}) because now $\nabla$
has a different meaning, see subsection~\ref{connection}.

Put $\nabla^\pm=\partial\pm iA$. Then the 2D Dirac equation is
\begin{equation}
\label{Dirac2D}
\!\!\!\!\!\!\!
\begin{pmatrix}
\ \ i\nabla^+_{\!0}-1\ \ &
\ \ 0\ \ &
\ \ 0\ \ &
\ \ i\nabla^+_{\!1}+\nabla^+_{\!2}\\
\ \ 0\ \ &
\ \ i\nabla^+_{\!0}-1\ \ &
\ \ i\nabla^+_{\!1}-\nabla^+_{\!2}\ \ &
\ \ 0\\
\ \ 0\ \ &
\ \ -i\nabla^+_{\!1}-\nabla^+_{\!2}\ \ &
\ \ -i\nabla^+_{\!0}-1\ \ &
\ \ 0\\
\ \ -i\nabla^+_{\!1}+\nabla^+_{\!2}\ \ &
\ \ 0\ \ &
\ \ 0\ \ &
\ \ -i\nabla^+_{\!0}-1
\end{pmatrix}
\begin{pmatrix}
\phi^{-+}\\\phi^{++}\\\chi^{++}\\\chi^{-+}
\end{pmatrix}
=0.
\end{equation}
Here we chose to use new notation for the components of the bispinor;
the relation with the traditional notation
is
$\psi=(\phi_1\ \phi_2\ \chi_1\ \chi_2)^T\!=
(\phi^{-+}\ \phi^{++}\ \chi^{++}\ \chi^{-+})^T\!$.

We shall also need the equation
\begin{equation}
\label{Dirac2Danti}
\!\!\!\!\!\!
\begin{pmatrix}
\ \ i\nabla^-_{\!0}-1\ \ &
\ \ 0\ \ &
\ \ 0\ \ &
\ \ i\nabla^-_{\!1}+\nabla^-_{\!2}\\
\ \ 0\ \ &
\ \ i\nabla^-_{\!0}-1\ \ &
\ \ i\nabla^-_{\!1}-\nabla^-_{\!2}\ \ &
\ \ 0\\
\ \ 0\ \ &
\ \ -i\nabla^-_{\!1}-\nabla^-_{\!2}\ \ &
\ \ -i\nabla^-_{\!0}-1\ \ &
\ \ 0\\
\ \ -i\nabla^-_{\!1}+\nabla^-_{\!2}\ \ &
\ \ 0\ \ &
\ \ 0\ \ &
\ \ -i\nabla^-_{\!0}-1
\end{pmatrix}
\begin{pmatrix}
\phi^{+-}\\\phi^{--}\\\chi^{--}\\\chi^{+-}
\end{pmatrix}
=0
\end{equation}
which is the 2D Dirac equation for the antiparticle; see formula
(32.5) in \cite{LL4}.

We shall write
$\nabla^\beta$, $\phi^{\alpha\beta}$, $\chi^{\alpha\beta}$
for $\nabla^\pm$, $\phi^{\pm\pm}$, $\chi^{\pm\pm}$,
and later $d^\beta$ for $d^\pm$.
Here we admit abusing
notation because $\alpha$ and $\beta$ were actually introduced
(see Sections~\ref{mass} and \ref{photon}) as numbers
and not signs.

The combined system (\ref{Dirac2Danti}), (\ref{Dirac2D})
can be written as
\begin{equation}
\label{combi}
\pmatrix
i\nabla^{\beta}_{\!0}-1\ \ &
i\nabla^{\beta}_{\!1}-\alpha\beta\nabla^{\beta}_{\!2}\\
-i\nabla^{\beta}_{\!1}-\alpha\beta\nabla^{\beta}_{\!2}\ \ &
-i\nabla^{\beta}_{\!0}-1
\endpmatrix
\pmatrix\phi^{\alpha\beta}\\\chi^{\alpha\beta}\endpmatrix=0.
\end{equation}

\begin{theorem}
\label{maintheorem}
The system (\ref{pertMax1}), (\ref{pertgauge1})
is equivalent to (\ref{combi}).
\end{theorem}

\noindent{\it Proof\ }
Using (\ref{genext3}) and (\ref{extder1}) we rewrite
(\ref{pertMax1}) as
\[
*(\partial\wedge v-*(A\wedge v))=
i\beta(\partial\wedge v-*(A\wedge v)).
\]
An elementary rearrangement of terms transforms the latter into
\[
*(\nabla^\beta\!\wedge v)=i\beta(\nabla^\beta\!\wedge v).
\]
Denoting $d^\beta Q:=\nabla^\beta\!\wedge Q$,
we see that (\ref{pertMax1}) takes the form
\begin{equation}
\label{proof1}
*d^\beta v=i\beta d^\beta v.
\end{equation}

Let us now write down explicitly our chosen photon $u$.
It is convenient to work in the coordinate system in which
the wave vector $k$ of our photon has components
$k_\mu=(1,0,0,\alpha)$;
this can always we achieved by a proper Lorentz transformation.
Straightforward calculations give
\begin{equation}
\label{proof2}
u_\mu=\pmatrix C\\1\\-i\alpha\beta\\ C\alpha\endpmatrix
e^{-i(x^0+\alpha x^3)}
\end{equation}
where $C$ is a constant (depending on the gauge).

Using (\ref{proof2}) it is easy to see that
$v$ satisfies (\ref{pertgauge1}) if and only if
\begin{equation}
\label{proof3}
v_\mu=
\left\{
\phi^{\alpha\beta}\pmatrix1\\0\\0\\\alpha\endpmatrix
-\chi^{\alpha\beta}\pmatrix0\\1\\ i\alpha\beta\\0\endpmatrix
\right\}e^{-i\alpha x^3}
\end{equation}
where $\phi^{\alpha\beta}$, $\chi^{\alpha\beta}$ are some
functions of $(x^0,x^1,x^2)$.

It remains to substitute (\ref{proof3}) into (\ref{proof1})
and obtain the equations for
$\phi^{\alpha\beta}$, $\chi^{\alpha\beta}$.
We have
\begin{multline}
\label{proof4}
\begin{pmatrix}
(*d^{\beta}v-i\beta d^{\beta}v)^{03}\\
(*d^{\beta}v-i\beta d^{\beta}v)^{13}\\
(*d^{\beta}v-i\beta d^{\beta}v)^{23}
\end{pmatrix}
=\begin{pmatrix}
-(d^{\beta}v)_{12}+i\beta(d^{\beta}v)_{03}\\
(d^{\beta}v)_{02}-i\beta(d^{\beta}v)_{13}\\
-(d^{\beta}v)_{01}-i\beta(d^{\beta}v)_{23}
\end{pmatrix}=
\\
\begin{pmatrix}
-i\beta\nabla^{\beta}_{\!3}\ &\nabla^{\beta}_{\!2}\ &
-\nabla^{\beta}_{\!1}\ &i\beta\nabla^{\beta}_{\!0}\\
-\nabla^{\beta}_{\!2}\ &i\beta\nabla^{\beta}_{\!3}\ &
\nabla^{\beta}_{\!0}\ &-i\beta\nabla^{\beta}_{\!1}\\
\nabla^{\beta}_{\!1}\ &-\nabla^{\beta}_{\!0}\ &
i\beta\nabla^{\beta}_{\!3}\ &-i\beta\nabla^{\beta}_{\!2}
\end{pmatrix}
\!\begin{pmatrix} v_0\\ v_1\\ v_2\\ v_3\end{pmatrix}
=\!\begin{pmatrix}
-\alpha\beta\ &\nabla^{\beta}_{\!2}\ &
-\nabla^{\beta}_{\!1}\ &i\beta\nabla^{\beta}_{\!0}\\
-\nabla^{\beta}_{\!2}\ &\alpha\beta\ &
\nabla^{\beta}_{\!0}\ &-i\beta\nabla^{\beta}_{\!1}\\
\nabla^{\beta}_{\!1}\ &-\nabla^{\beta}_{\!0}\ &
\alpha\beta\ &-i\beta\nabla^{\beta}_{\!2}
\end{pmatrix}
\!\begin{pmatrix} v_0\\ v_1\\ v_2\\ v_3\end{pmatrix}
\\
=\begin{pmatrix}
\alpha\beta(i\nabla^{\beta}_{\!0}-1)\ \ &
i\alpha\beta\nabla^{\beta}_{\!1}-\nabla^{\beta}_{\!2}\\
-i\alpha\beta\nabla^{\beta}_{\!1}-\nabla^{\beta}_{\!2}\ \ &
-i\alpha\beta\nabla^{\beta}_{\!0}-\alpha\beta\\
\nabla^{\beta}_{\!1}-i\alpha\beta\nabla^{\beta}_{\!2}\ \ &
\nabla^{\beta}_{\!0}-i
\end{pmatrix}
\begin{pmatrix}
\phi^{\alpha\beta}\\\chi^{\alpha\beta}
\end{pmatrix} e^{-i\alpha x^3}.
\end{multline}
The last two lines of the matrix in the right hand side of
(\ref{proof4}) are linearly dependent, so (\ref{proof1})
reduces to
\[
\begin{pmatrix}
\alpha\beta(i\nabla^{\beta}_{\!0}-1)\ \ &
i\alpha\beta\nabla^{\beta}_{\!1}-\nabla^{\beta}_{\!2}\\
-i\alpha\beta\nabla^{\beta}_{\!1}-\nabla^{\beta}_{\!2}\ \ &
-i\alpha\beta\nabla^{\beta}_{\!0}-\alpha\beta
\end{pmatrix}
\begin{pmatrix}\phi^{\alpha\beta}\\\chi^{\alpha\beta}\end{pmatrix}=0.
\]
Multiplying the latter by $\alpha\beta$ we arrive at (\ref{combi}).

\section{Free particles}
\label{free}

Let us consider the situation when there is no external electromagnetic
field, i.e., $A\equiv0$. In this case our
system (\ref{pertMax1}), (\ref{pertgauge1}) has a variety of plane wave
solutions, out of which we single out one particular in accordance
with the following

\begin{assumption}
\label{free1}
The only physically meaningful plane wave
solution $v$ is the one whose wave vector is the same as for $u$.
\end{assumption}

This physical assumption is made in the spirit of Feynman
diagrams. One would expect that on the basis of
(\ref{Maxwell3}), (\ref{pertMax1}), (\ref{pertgauge1})
it would be possible to develop
a full perturbation theory describing the interaction of
electrons, positrons and photons
(tensor analogue of Feynman diagrams),
and the above physical assumption would emerge as a natural
consequence of this theory.
In its absence we have to content ourselves with
introducing Physical Assumption \ref{free1} as an axiom.

Up to a proper Lorentz transformation and complex conjugation
(see Section~\ref{basic})
all our physically meaningful plane wave solutions
can be written as
\begin{equation}
\label{free2}
u_\mu=\pmatrix C\\1\\-i\alpha\beta\\ C\alpha\endpmatrix
e^{-i(x^0+\alpha x^3)},
\qquad
v_\mu=\pmatrix1\\0\\0\\\alpha\endpmatrix
e^{-i(x^0+\alpha x^3)}.
\end{equation}
Here, as in (\ref{proof2}), $C$ is an arbitrary constant.

We shall call the vector function $v$ in (\ref{free2}) the
free electron/positron.
We see that the free electron/positron is
a gradient type solution of the polarised Maxwell equation.

\section{Distinguishing the electron and the positron}
\label{charge}

Let us now separate the plane wave solutions (\ref{free2})
corresponding to the electron and the positron. As we already
have Theorem~\ref{maintheorem} and formula (\ref{proof3}),
the separation procedure reduces to the
analysis of the case of a weak constant purely electric vector
potential $A$. Namely, we look for plane wave solutions $v$ of
the form
\[
\text{constant vector}
\ \times\ e^{-i(\epsilon x^0+\alpha x^3)}
\]
which are perturbations of (\ref{free2}), i.e., $\epsilon\approx1$.
We say that we are dealing with an
electron if $\epsilon=1+A_0$, and with a positron if
$\epsilon=1-A_0$. As a result we arrive at the following
classification of plane wave solutions (\ref{free2}):
solutions
\begin{equation}
\label{charge1}
u_\mu=\pmatrix C\\1\\-i\alpha\\ C\alpha\endpmatrix
e^{-i(x^0+\alpha x^3)},
\qquad
v_\mu=\pmatrix1\\0\\0\\\alpha\endpmatrix
e^{-i(x^0+\alpha x^3)}
\end{equation}
correspond to the free electron, whereas solutions
\begin{equation}
\label{charge2}
u_\mu=\pmatrix C\\1\\ i\alpha\\ C\alpha\endpmatrix
e^{-i(x^0+\alpha x^3)},
\qquad
v_\mu=\pmatrix1\\0\\0\\\alpha\endpmatrix
e^{-i(x^0+\alpha x^3)}
\end{equation}
correspond to the free positron. As usual, formulae
(\ref{charge1}), (\ref{charge2}) are written up to a proper
Lorentz transformation and complex conjugation.

Comparing (\ref{charge1}) with (\ref{charge2}) we see that looking
only at the vector function $v$ it is impossible to distinguish
the free electron from the free positron: the difference occurs
in the formulae for the associated photon $u$. Physically
this means that it is impossible to tell whether we are dealing
with an electron or a positron until we examine how the particle
interacts with the electromagnetic field.

A convenient way of distinguishing the two cases is to define
the notion of charge in accordance with
\begin{equation}
\label{charge3}
{\mathbf c}:=-\operatorname{sgn}
\left(
i*\left(du\wedge\overline{{\mathcal R}du}\,\right)
\right)
\end{equation}
where ${\mathcal R}$ is the reflection operator (\ref{reflection}).
Substituting (\ref{charge1}) and (\ref{charge2}) into
(\ref{charge3}) and performing straightforward calculations
we conclude that for the electron ${\mathbf c}=-1$,
whereas for the positron ${\mathbf c}=+1$.

It is easy to see that ${\mathbf c}$ is a true scalar in that it
is invariant under Lorentz transformations (proper and improper)
and does not depend on the choice of gauge for $u$.
Moreover, at a formal mathematical level our definition of charge
(\ref{charge3}) works in the case of an external field $A$, and
even irrespective of the strength of this field.

On the other hand ${\mathbf c}$ is not invariant under
the transformation $u\to\overline u$. This means that
the notion of charge can only be used if we distinguish
the forward and backward light cones, i.e., specify
the positive direction of time.

\end{document}